\documentclass[conference]{IEEEtran}
\IEEEoverridecommandlockouts
\usepackage{cite}
\usepackage{amsmath,amssymb,amsfonts}
\usepackage{algorithmic}
\usepackage{graphicx}
\usepackage{makecell}
\usepackage{textcomp}
\usepackage{xcolor}
\usepackage[colorlinks=true, linkcolor=black, citecolor=black, urlcolor=black]{hyperref}
\def\BibTeX{{\rm B\kern-.05em{\sc i\kern-.025em b}\kern-.08em
    T\kern-.1667em\lower.7ex\hbox{E}\kern-.125emX}}
    
\begin{document}

\title{Content-based Recommendation Engine for Video Streaming Platform}

\author{
     \IEEEauthorblockN{Puskal Khadka}
    \IEEEauthorblockA{
        \textit{Department of Computer Science} \\
        \textit{Tribhuvan University} \\
         Kathmandu, Nepal \\
        puskalkumar.khadka@texasintl.edu.np
    }
    \and
    \IEEEauthorblockN{Prabhav Lamichhane}
    \IEEEauthorblockA{
        \textit{Department of Computer Science} \\
        \textit{Tribhuvan University} \\
        Kathmandu, Nepal \\
        prabhav.lamichhane@texasintl.edu.np
    }   
}

\maketitle

\begin{abstract}
\hspace{4pt}Recommendation engines suggest content, products, or services to the user by using machine learning algorithms. This paper proposes a content-based recommendation engine that provides personalized video suggestions based on users’ previous interactions and preferences. The engine uses TF-IDF (Term Frequency-Inverse Document Frequency) text vectorization technique to evaluate the relevance of words in video descriptions, followed by the computation of cosine similarity between content items to determine their degree of similarity. The system’s performance is evaluated using precision, recall, and F1-score metrics. Experimental results demonstrate the effectiveness of content-based filtering in delivering relevant and personalized video recommendations to users. This approach can enhance user engagement on video streaming platforms and reduce search time, providing a more intuitive, preference-based viewing experience. 
\end{abstract}
\vspace{4pt}

\section{Introduction}
Video streaming platforms have drastically increased in recent years due to technological advancement and the widespread availability of handheld devices. Each platform contains a huge amount of content across different genres, giving users a wide range of choices. However, users are often perplexed by the overwhelming amount of content available on these platforms. As the volume of digital media continues to grow rapidly, manually searching for relevant items becomes increasingly difficult and time-consuming. Therefore, it is necessary to implement an effective recommendation engine. These systems assist users by filtering and identifying relevant items from massive datasets and also help platforms to retain users by presenting content that aligns with their interests.

There are mainly three types of recommendation systems:  content-based filtering, collaborative filtering, and hybrid recommendation systems~\cite{adomavicius2005toward}. Each of these systems uses different techniques to provide personalized recommendations to end users.

Content-based filtering suggests results to users based on their previously preferred interests and choices~\cite{Balabanovi1997fab, Pazzani2007content}. It uses machine learning techniques to learn the similarity between different items and makes predictions accordingly. This approach generally creates a user profile based on the types of items the user likes and then recommends new items by comparing their features with the generated profile~\cite{lops2011contentbased}. Content-based filtering performs well even when user–user interaction data is limited as it relies solely on item metadata such as genre, cast, keywords, or descriptions. Additionally, it can continuously refine its recommendations as users interact with more items which makes it stable and adaptable.

Collaborative filtering systems collect and analyze data on users’ behavior and predict results based on similarities with other users~\cite{sarwar2001itembased, zhang2014collaborative}. Instead of requiring prior knowledge of item features, it operates under the assumption that users may like items preferred by other users with similar preferences. Despite its popularity and wide adoption in platforms such as Amazon and Netflix, collaborative filtering suffers from the cold-start problem~\cite{ijcai2021preference} and sparsity issue~\cite{grcar2006datasparsity}. Some early studies have addressed this problem using matrix factorization in collaborative filtering~\cite{koren2009matrix, ocepek2015improving}.

Hybrid recommendation systems~\cite{ccano2017hybrid,trabelsi2021hybridrs} combine both collaborative and content-based filtering. This approach significantly reduces the weaknesses of individual filtering methods by leveraging both a user’s past preferences and similarities with other users who have related interests~\cite{ricci2015recommender}. As a result, hybrid systems often achieve higher accuracy, greater diversity in suggestions, and stronger robustness when applying to various applications.

In this paper, we present a content-based filtering approach for designing a recommendation engine that predicts users’ interests based on features from previously watched content, such as cast, genre, and overview. By focusing on item attributes, the proposed system ensures consistent performance even when user behavior data is limited and makes it more suitable for platforms with many new users or frequently updated content libraries.

\section{Problem Statement}
Video streaming platforms generally comprise a large amount of video content. In most cases, users are generally inclined towards a particular genre, and there is a high tendency that users will consume similar types of content in the future. But without a recommendation engine on the platform, if the user wants to stream content of their own choices and preferences, each time they mostly have to manually search or filter content from the large pool of available videos. This process is generally tedious and repetitive, especially as the size of the content library continues to grow. As a result, the overall user experience becomes less efficient and less engaging. Therefore, to solve this problem, there is a need for an appropriate recommendation engine on the platform that provides accurate suggestions to users based on their past watching choices and habits.

\section{Related Work}
Different types of recommendation engines have been researched and developed over the last few decades~\cite{lu2015recommender}. They are widely adopted on several platforms such as Amazon~\cite{linden2003amazon}, Netflix~\cite{uribe2016netflix}, Facebook~\cite{gupta2020architectural}, TikTok~\cite{liu2022monolith} and YouTube~\cite{Covington2016DeepNN} to enhance user engagement by providing personalized content and improving user retention. Beside this, multiple studies have been conducted over the decades to find the effective recommender system~\cite{zhang2019deep, Lops2011content, thorat2015survey}.

Amazon's recommender system~\cite{linden2003amazon} uses item-to-item collaborative filtering, which predicts user interest based on visitors’ recent purchase history, browsing behavior, and ratings. This method effectively identifies patterns in user behavior and generates recommendations by finding similarities between items purchased or viewed by similar users. However, collaborative filtering may face challenges such as the cold-start problem, where new items or users lack sufficient data for accurate predictions.

Baatarjav et al.~\cite{Baatarjav2008group} introduces a group recommendation system (GRS) for Facebook using a combination of hierarchical clustering and decision tree techniques. The system analyzes the preferences of multiple users to provide recommendations that satisfy group interests. Based on experimental results, the authors report that the GRS achieves improved accuracy, highlighting the importance of extracting relevant item-level features and combining them effectively to enhance prediction performance.

Shani and Gunawardana~\cite{Shani2011evaluating} evaluate recommendation systems by outlining experimental settings appropriate for comparing different algorithms. They discuss three experimental settings: offline evaluations that compare recommendation methods without user interaction, user studies that examine user experience, and large-scale online experiments that analyze system performance with real users. Their study emphasizes the need for systematic evaluation of recommendation systems to ensure generalizability of the results.

Overall, these studies underline the importance of feature extraction, accurate similarity measurement, and rigorous evaluation in designing effective recommendation engines. While collaborative and group-based approaches are widely used, content-based filtering remains particularly effective when user-item interaction data is sparse or when recommendations need to be personalized based on item characteristics.

\section{Methodology}
In this section, we present the detailed methodology for designing the proposed video recommender system. First, we collected datasets representing diverse videos. Next, the raw datasets are preprocessed and cleaned to ensure consistency and high data quality. Following preprocessing, an appropriate text vectorization technique is applied to quantify the significance of each term within the corpus. Finally, a similarity measurement algorithm is used to compute pairwise similarities between videos, using which the recommendation is generated.

\subsection{Data Gathering and Preprocessing}
The efficiency of a recommendation system is highly dependent on the quantity and quality of available data. For this study, we used a movie dataset~\cite{kaggle2023movie} consisting of 4,803 movies, which serves as the video content for our proposed streaming platform. For demonstration purposes, we selected five movies from the original dataset to illustrate the internal computations and functioning of our recommendation engine. Each movie includes multiple attributes, such as genre, original language, release date, cast, overview, and production company. From these, we selected three primary features, Genre, Cast, and Overview and compiled them into a separate table, as shown in Table~\ref{table:movie_dataset}.
 \begin{table}[htbp]
\caption{Sample Movie/Video Datasets}
\centering
\renewcommand{\arraystretch}{1.3} 
\resizebox{0.49\textwidth}{!}{
\begin{tabular}{|l|l|l|l|}
\hline
\textbf{Movie} & \textbf{Genre} & \textbf{Cast} & \textbf{Overview} \\
\hline
Ironman & scifi & Robert Downey Jr & mcu, weaponed suit, super hero \\
\hline
Titanic & romance & Leonardo Dicaprio & sea disaster, romance \\
\hline
Avengers & scifi & Robert Downey Jr, chris evans & mcu, shield, super hero \\
\hline
Great Gatsby & romance, novel & Leonardo Dicaprio & social difference, obsession, novel \\
\hline
Forrest Gump & novel & Tom Hank & inspirational, low iq, romance \\
\hline
\end{tabular}
}
\label{table:movie_dataset}
\end{table}

After creating the sample dataset, several data cleaning and preprocessing steps were performed, including the removal of stop words, null values, and special characters from each document. Then, data from the three selected columns for each movie were combined into a single filtered document corpus, as given below.
\\ \\
\textbf{Ironman}: scifi robertdowneyjr mcu weaponedsuit superhero\\
\textbf{Titanic:} romance leonardodicarpio seadisaster romance\\
\textbf{Avengers:} scifi robertdowneyjr chrisevans mcu shield superhero\\
\textbf{Great gatsby:} romance novel leonardodicarpio socialdifference
obsession novel\\
\textbf{Forrest gump:} novel tomhank inspirational romance

\subsection{Text Vectorization}
After getting the filtered dataset, the corpus is converted into a vector representation. These vectors play a crucial role in constructing the recommendation model. Several text vectorization techniques~\cite{yang2022vectorization, kumari2019vectorization} exist in natural language processing, including Bag of Words, TF-IDF, Word2Vec, and so on. In this work, we use the TF-IDF method to convert the filtered corpus into vectors. TF-IDF identifies the most relevant and unique terms in each document and is computed using two components: TF and IDF.

\subsubsection{Term Frequency (TF)}
Term frequency finds out the frequency of appearance of a particular term in the given document. For a document $d$, the term frequency of a word $w$ is defined as:
\begin{equation}
    \text{TF}(w,d) = \frac{\text{Number of times word } w \text{ appears in } d}{\text{Total number of words in } d} \nonumber
\end{equation}
Using this formula, the TF for each word in the corpus is calculated, as shown in the Table~\ref{table:term_frequency}.
\begin{table}[htbp]
\caption{Term Frequency of Word in Corpus}
\centering
\renewcommand{\arraystretch}{1.3} 
\resizebox{\linewidth}{!}{
\begin{tabular}{l|c|c|c|c|c}
\hline
\textbf{} & \textbf{Ironman} & \textbf{Titanic} & \textbf{Avengers} & \textbf{Great Gatsby} & \textbf{Forrest Gump} \\
\hline
\textbf{scifi} & 1/5 & 0/4 & 1/6 & 0/6 & 0/4 \\
\hline
\textbf{robertdowneyjr} & 1/5 & 0/4 & 1/6 & 0/6 & 0/4 \\
\hline
\textbf{mcu} & 1/5 & 0/4 & 1/6 & 0/6 & 0/4 \\ 
\hline
\textbf{weaponedsuit} & 1/5 & 0/4 & 0/6 & 0/6 & 0/4 \\
\hline
\textbf{superhero} & 1/5 & 0/4 & 1/6 & 0/6 & 0/4 \\
\hline
\textbf{romance} & 0/5 & 2/4 & 0/6 & 1/6 & 1/4 \\
\hline
\textbf{leonardodicaprio} & 0/5 & 1/4 & 0/6 & 1/6 & 0/4 \\
\hline
\textbf{seadisaster} & 0/5 & 1/4 & 0/6 & 0/6 & 0/4 \\
\hline
\textbf{chrisevans} & 0/5 & 0/4 & 1/6 & 0/6 & 0/4 \\
\hline
\textbf{shield} & 0/5 & 0/4 & 1/6 & 0/6 & 0/4 \\
\hline
\textbf{novel} & 0/5 & 0/4 & 0/6 & 2/6 & 1/4 \\
\hline
\textbf{socialdifference} & 0/5 & 0/4 & 0/6 & 1/6 & 0/4 \\
\hline
\textbf{obsession} & 0/5 & 0/4 & 0/6 & 1/6 & 0/4 \\
\hline
\textbf{tomhank} & 0/5 & 0/4 & 0/6 & 0/6 & 1/4 \\
\hline
\textbf{inspirational} & 0/5 & 0/4 & 0/6 & 0/6 & 1/4 \\
\hline
\end{tabular}
}
\label{table:term_frequency}
\end{table}

\subsubsection{Inverse Document Frequency (IDF)}
While a high term frequency (TF) suggests that a word appears frequently in a document, it does not always indicate its importance for video recommendation. Common words, such as ``movie'' may appear multiple times but carry little significance in distinguishing content. To address this limitation, the inverse document frequency (IDF) is calculated, which reduces the weight of frequently occurring and less informative words across the documents.
\begin{equation}
\text{IDF}(w) = \log \frac{\text{Total number of videos}}{\text{Number of videos containing word } w} \nonumber
\end{equation}
\begin{table}[h!]
\centering
\caption{IDF of Each Word in Corpus}
\renewcommand{\arraystretch}{1.1}  
\resizebox{0.48\linewidth}{!}{
\begin{tabular}{l|l}
\hline
\textbf{} & \textbf{IDF} \\
\hline
\textbf{scifi} & $\log(5/2)$ \\
\hline
\textbf{robertdowneyjr} & $\log(5/2)$ \\
\hline
\textbf{mcu} & $\log(5/2)$ \\
\hline
\textbf{weaponedsuit} & $\log(5/1)$ \\
\hline
\textbf{superhero} & $\log(5/2)$ \\
\hline
\textbf{romance} & $\log(5/3)$ \\
\hline
\textbf{leonardodicaprio} & $\log(5/2)$ \\
\hline
\textbf{seadisaster} & $\log(5/1)$ \\
\hline
\textbf{chrisevans} & $\log(5/1)$ \\
\hline
\textbf{shield} & $\log(5/1)$ \\
\hline
\textbf{novel} & $\log(5/2)$ \\
\hline
\textbf{socialdifference} & $\log(5/1)$ \\
\hline
\textbf{obsession} & $\log(5/1)$ \\
\hline
\textbf{tomhank} & $\log(5/1)$ \\
\hline
\textbf{inspirational} & $\log(5/1)$ \\
\hline
\end{tabular}
}
\label{table:idf}
\end{table}

The logarithm is used to prevent large IDF values from dominating TF. The resulting IDF values for the corpus are presented in the Table~\ref{table:idf}.

\subsubsection{TF-IDF}
A word in a document is considered most important when it occurs frequently within that document but rarely in other documents. Since TF captures the frequency of a word in its own document and IDF captures its rarity across other documents, the actual relevance of a word is obtained by multiplying these two measures. Tables~\ref{table:term_frequency} and~\ref{table:idf} present the TF and IDF values, respectively, which are then combined to compute the TF-IDF scores.
\begin{equation}
    \text{TF-IDF}(w,d) = \text{TF}(w,d) \times \text{IDF}(w) \nonumber
\end{equation}
TF-IDF value for each word is shown in the following
tables where each cell shows the multiplication of TF and IDF
value.

\begin{table}[htbp]
\caption{TF-IDF Value for Each Word}
\centering
\renewcommand{\arraystretch}{2.5} 
\resizebox{\linewidth}{!}{
\begin{tabular}{l|c|c|c|c|c}
\hline
\textbf{} & \textbf{scifi} & \textbf{robertdowneyjr} & \textbf{mcu} & \textbf{weaponedsuit} & \textbf{superhero} \\
\hline
\textbf{Ironman} & \makecell[l]{(1/5)*log(5/2)\\= \textbf{0.0795880}} & \makecell[l]{(1/5)*log(5/2)\\= \textbf{0.0795880}} & \makecell[l]{(1/5)*log(5/2)\\= \textbf{0.0795880}} & \makecell[l]{(1/5)*log(5/1)\\= \textbf{0.1397940}} & \makecell[l]{(1/5)*log(5/2)\\= \textbf{0.0795880}} \\
\hline
\textbf{Titanic} & \makecell[l]{(0/4)*log(5/2)\\= \textbf{0}} & \makecell[l]{(0/4)*log(5/2)\\= \textbf{0}} & \makecell[l]{(0/4)*log(5/2)\\= \textbf{0}} & \makecell[l]{(0/4)*log(5/1)\\= \textbf{0}} & \makecell[l]{(0/4)*log(5/2)\\= \textbf{0}} \\
\hline
\textbf{Avengers} & \makecell[l]{(1/6)*log(5/2)\\= \textbf{0.0663233}} & \makecell[l]{(1/6)*log(5/2)\\= \textbf{0.0663233}} & \makecell[l]{(1/6)*log(5/2)\\= \textbf{0.0663233}} & \makecell[l]{(0/6)*log(5/1)\\= \textbf{0}} & \makecell[l]{(1/6)*log(5/2)\\= \textbf{0.0663233}} \\
\hline
\textbf{Great gatsby} & \makecell[l]{(0/6)*log(5/2)\\= \textbf{0}} & \makecell[l]{(0/6)*log(5/2)\\= \textbf{0}} & \makecell[l]{(0/6)*log(5/2)\\= \textbf{0}} & \makecell[l]{(0/6)*log(5/1)\\= \textbf{0}} & \makecell[l]{(0/6)*log(5/2)\\= \textbf{0}} \\
\hline
\textbf{Forrest gump} & \makecell[l]{(0/4)*log(5/2)\\= \textbf{0}} & \makecell[l]{(0/4)*log(5/2)\\= \textbf{0}} & \makecell[l]{(0/4)*log(5/2)\\= \textbf{0}} & \makecell[l]{(0/4)*log(5/1)\\= \textbf{0}} & \makecell[l]{(0/4)*log(5/2)\\= \textbf{0}} \\
\hline
\end{tabular}
}
\label{table:tfidf1}
\end{table}

\begin{table}[htbp]
\caption{TF-IDF Value for Each Word (Continued-1)}
\centering
\renewcommand{\arraystretch}{2.5}  
\resizebox{\linewidth}{!}{
\begin{tabular}{l|c|c|c|c|c}
\hline
\textbf{} & \textbf{romance} & \textbf{leonardodicaprio} & \textbf{seadisaster} & \textbf{chrisevans} & \textbf{shield} \\
\hline
\textbf{Ironman} & \makecell[l]{(0/5)*log(5/3)\\= \textbf{0}} & \makecell[l]{(0/5)*log(5/2)\\= \textbf{0}} & \makecell[l]{(0/5)*log(5/1)\\= \textbf{0}} & \makecell[l]{(0/5)*log(5/1)\\= \textbf{0}} & \makecell[l]{(0/5)*log(5/1)\\= \textbf{0}} \\
\hline
\textbf{Titanic} & \makecell[l]{(2/4)*log(5/3)\\= \textbf{0.1109244}} & \makecell[l]{(1/4)*log(5/2)\\= \textbf{0.0994850}} & \makecell[l]{(1/4)*log(5/1)\\= \textbf{0.1747425}} & \makecell[l]{(0/4)*log(5/1)\\= \textbf{0}} & \makecell[l]{(0/4)*log(5/1)\\= \textbf{0}} \\
\hline
\textbf{Avengers} & \makecell[l]{(0/6)*log(5/3)\\= \textbf{0}} & \makecell[l]{(0/6)*log(5/2)\\= \textbf{0}} & \makecell[l]{(0/6)*log(5/1)\\= \textbf{0}} & \makecell[l]{(1/6)*log(5/1)\\= \textbf{0.1164950}} & \makecell[l]{(1/6)*log(5/1)\\= \textbf{0.1164950}} \\
\hline
\textbf{Great gatsby} & \makecell[l]{(1/6)*log(5/3)\\= \textbf{0.0369748}} & \makecell[l]{(1/6)*log(5/2)\\= \textbf{0.0663233}} & \makecell[l]{(0/6)*log(5/1)\\= \textbf{0}} & \makecell[l]{(0/6)*log(5/1)\\= \textbf{0}} & \makecell[l]{(0/6)*log(5/1)\\= \textbf{0}} \\
\hline
\textbf{Forrest gump} & \makecell[l]{(1/4)*log(5/3)\\= \textbf{0.0554622}} & \makecell[l]{(0/4)*log(5/2)\\= \textbf{0}} & \makecell[l]{(0/4)*log(5/1)\\= \textbf{0}} & \makecell[l]{(0/4)*log(5/1)\\= \textbf{0}} & \makecell[l]{(0/4)*log(5/1)\\= \textbf{0}} \\
\hline
\end{tabular}
}
\label{table:tfidf2}
\end{table}

\begin{table}[htbp]
\caption{TF-IDF Value for Each Word (Continued-2)}
\renewcommand{\arraystretch}{2.5}
\centering
\resizebox{\linewidth}{!}{
\begin{tabular}{l|c|c|c|c|c}
\hline
\textbf{} & \textbf{novel} & \textbf{socialdifference} & \textbf{obsession} & \textbf{tomhank} & \textbf{inspirational} \\
\hline
\textbf{Ironman} & \makecell[l]{(0/5)*log(5/2)\\= \textbf{0}} & \makecell[l]{(0/5)*log(5/1)\\= \textbf{0}} & \makecell[l]{(0/5)*log(5/1)\\= \textbf{0}} & \makecell[l]{(0/5)*log(5/1)\\= \textbf{0}} & \makecell[l]{(0/5)*log(5/1)\\= \textbf{0}} \\
\hline
\textbf{Titanic} & \makecell[l]{(0/4)*log(5/2)\\= \textbf{0}} & \makecell[l]{(0/4)*log(5/1)\\= \textbf{0}} & \makecell[l]{(0/4)*log(5/1)\\= \textbf{0}} & \makecell[l]{(0/4)*log(5/1)\\= \textbf{0}} & \makecell[l]{(0/4)*log(5/1)\\= \textbf{0}} \\
\hline
\textbf{Avengers} & \makecell[l]{(0/6)*log(5/2)\\= \textbf{0}} & \makecell[l]{(0/6)*log(5/1)\\= \textbf{0}} & \makecell[l]{(0/6)*log(5/1)\\= \textbf{0}} & \makecell[l]{(0/6)*log(5/1)\\= \textbf{0}} & \makecell[l]{(0/6)*log(5/1)\\= \textbf{0}} \\
\hline
\textbf{Great gatsby} & \makecell[l]{(2/6)*log(5/2)\\= \textbf{0.1326467}} & \makecell[l]{(1/6)*log(5/1)\\= \textbf{0.1164950}} & \makecell[l]{(1/6)*log(5/1)\\= \textbf{0.1164950}} & \makecell[l]{(0/6)*log(5/1)\\= \textbf{0}} & \makecell[l]{(0/6)*log(5/1)\\= \textbf{0}} \\
\hline
\textbf{Forrest gump} & \makecell[l]{(1/4)*log(5/2)\\= \textbf{0.0994850}} & \makecell[l]{(0/4)*log(5/1)\\= \textbf{0}} & \makecell[l]{(0/4)*log(5/1)\\= \textbf{0}} & \makecell[l]{(1/4)*log(5/1)\\= \textbf{0.1747425}} & \makecell[l]{(1/4)*log(5/1)\\= \textbf{0.1747425}} \\
\hline
\end{tabular}
}
\label{table:tfidf3}
\end{table}

\vspace{60pt}
Value in each cell in the above table represents the importance of a word in the document, and each row forms a sequence of numbers. This ordered sequence is referred to as a numeric vector, which can be used to compute similarity between documents~\cite{wael2013survey}. The final combined sequence of values from each row of the TF-IDF tables, which will represents the numeric vector for each movie, is presented in the Table~\ref{table:movievectors}.

\begin{table}[htbp!]
\caption{Vector for Each Movie}
\centering
\renewcommand{\arraystretch}{2} 
\resizebox{\linewidth}{!}{
\begin{tabular}{l|l}
\hline
\textbf{Ironman} & ( 0.0795880, 0.0795880, 0.0795880, 0.1397940, 0.0795880, 0, 0, 0, 0, 0, 0, 0, 0, 0, 0 ) \\
\hline
\textbf{Titanic} & ( 0, 0, 0, 0, 0, 0.1109244, 0.0994850, 0.1747425, 0, 0, 0, 0, 0, 0, 0 ) \\
\hline
\textbf{Avengers} & (  0.0663233, 0.0663233, 0.0663233, 0, 0.0663233, 0, 0, 0, 0.1164950, 0.1164950, 0, 0, 0, 0, 0 ) \\
\hline
\textbf{Great gatsby} & ( 0, 0, 0, 0, 0, 0.0369748, 0.0663233, 0, 0, 0, 0.1326467, 0.1164950, 0.1164950, 0, 0 ) \\
\hline
\textbf{Forrest gump} & ( 0, 0, 0, 0, 0, 0.0554622, 0, 0, 0, 0, 0.0994850, 0, 0, 0.1747425,  0.1747425 ) \\
\hline
\end{tabular}
}
\label{table:movievectors}
\end{table}

\subsubsection{Cosine similarity}
Cosine similarity measures the similarity between two vectors by computing the cosine of the angle between them~\cite{victor2014csmr}. Table~\ref{table:movievectors} shows the corresponding vectors for each movie. Mathematically, the cosine similarity between these two movies can be expressed as:

\begin{equation}
    \cos(\theta) = \frac{\mathbf{A} \cdot \mathbf{B}}{\|\mathbf{A}\| \|\mathbf{B}\|} = 
\frac{\displaystyle\sum_{i=1}^{n} A_i B_i}
     {\sqrt{\displaystyle\sum_{i=1}^{n} A_i^2} \sqrt{\displaystyle\sum_{i=1}^{n} B_i^2}} \nonumber
\end{equation}

where $A_{i}$ and $B_{i}$ are $i^{th}$ components of two vectors A and B, respectively.

\begin{figure}[htbp]
    \centering
    \includegraphics[width=\linewidth]{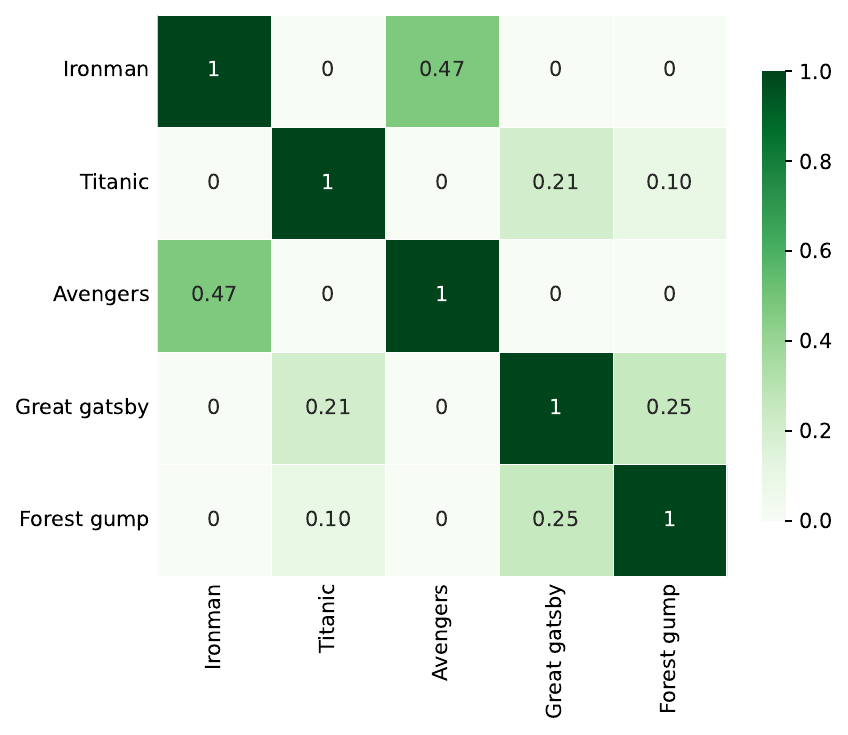}
    \caption{Heatmap showing cosine similarity between movies}
    \label{fig:heatmap}
\end{figure}

By using the above formula, we calculate cosine similarity between every vector and obtained results which are plotted in the heatmap, as shown in Fig.~\ref{fig:heatmap}.

The range of cosine similarity lies between -1 and 1. However, since the TF-IDF vectors contain only non-negative values, the cosine similarity scores obtained in our case range between 0 and 1. Value 1 means that two documents are exactly similar to each other, whereas 0 means that the two documents are completely dissimilar.

Once the similarity scores between movies are calculated, as shown in Fig.~\ref{fig:heatmap}, the recommendation engine uses these scores to suggest relevant videos. Before generating recommendations, the system first examines the user’s past viewing history and interests. During the auto-suggestion process, the engine evaluates the cosine similarity values for the movies that are previously watched or searched and recommends those that have the maximum similarity score.

Suppose a user previously watched the movie ‘Ironman’ on our video streaming platform. The next time, while suggesting, our system will look only for those movies, which have high cosine similarity with the movie ‘Ironman’. If we look at our heatmap in the Fig.~\ref{fig:heatmap}, ‘Avengers’ has the highest cosine similarity value with “Ironman’ which is 0.47. So next time, our system will auto-recommend the movie ‘Avengers’ to the user. But it will not recommend movies such as ‘Titanic’, ‘Great gatsby’ because if we look at the cosine similarity table in the above heatmap, the similarity score of movies ‘Titanic’, ‘Great gatsby’ with the movie ‘Ironman’ is zero.

\section{Results and Analysis}
We evaluated the recommendation engine using a user-centric approach. Since content-based filtering focuses on modeling individual user preferences, the evaluation is conducted from the perspective of a user. We select a user, i.e, User A, and construct a separate analytical profile capturing his interactions, viewing preferences, and engagement patterns with the video content available on the platform. Using this profile as input, the recommender system generates a predicted set of videos that the user may find relevant. Finally, we compare the predicted recommendations with the user’s actual preferences to identify the four possible categories, which are collectively represented in the form of a confusion matrix, as shown in Table~\ref{table:confusion_matrix}.
\begin{table}[h]
\centering
\caption{Confusion Matrix of Our Recommender}
\renewcommand{\arraystretch}{1.5} %
\resizebox{\linewidth}{!}{
\begin{tabular}{c|c|c}
\hline
\textbf{} & \textbf{Recommended} & \textbf{Not Recommended} \\
\hline
\textbf{Interested} & True Positive (TP): 10 & False Negative (FN): 2 \\
\hline
\textbf{Not Interested} & False Positive (FP): 1 & True Negative (TN): 4 \\
\hline
\end{tabular}
\label{table:confusion_matrix}
}
\end{table}
\\
From the above table, it can be seen that our recommendation engine produces a total of 17 predictions for User~A. Out of these, the system correctly recommends 10 movies in which the user is actually interested, representing the True Positive (TP) count. Similarly, the number of items in which User~A is interested but that are not recommended by the system is 2, which corresponds to the False Negative (FN) value. There is also one item that the user is not interested in but is still recommended by the system, representing the False Positive (FP) count. Finally, the total number of items that the user is not interested in and that are also not recommended by the system is 4, representing the True Negative (TN) count. Using this confusion matrix value, we compute the performance metrics of our recommendation engine in the following subsections.

\subsection{Precision}
Precision gives an indication of how many of the items recommended by the system are actually relevant to the user. It shows the quality and reliability of the system’s positive predictions. Mathematically, precision is defined as:
\begin{equation}
\text{Precision} = \frac{TP}{TP + FP} \nonumber
\end{equation}

For User~A, the recommendation engine correctly identifies 10 relevant items (TP) while making only 1 incorrect recommendation (FP). Using these values, we get the \textbf{precision of the system as 0.909}. This means that nearly 91\% of the videos recommended to the user align with the actual viewing preferences of that user.

\subsection{Recall}

Recall measures the proportion of all relevant items that are successfully recommended to the user. It indicates the system’s ability to capture user interest correctly. Recall can be mathematically defined as:
\begin{equation}
\text{Recall} = \frac{TP}{TP + FN} \nonumber
\end{equation}

In the case of User~A, the system retrieves 10 of the 12 items the user is interested in, resulting in a \textbf{recall value of 0.833}. This shows that our recommendation engine is able to identify a substantial portion of the user’s preferred content, though some relevant items are still missed.

\begin{figure}[t]
    \centering
    \includegraphics[width=0.98\linewidth]{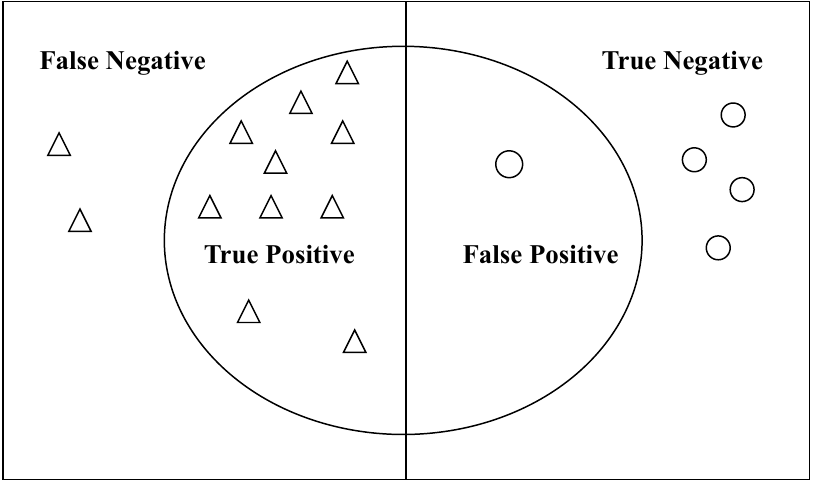}
    \caption{Venn diagram illustrating the relationship between precision and recall}
    \label{fig:fptp}
\end{figure}

\subsection{F1 Score}
The F1 score combines precision and recall into a single performance metric by taking their harmonic mean. This metric is useful for evaluating systems where both the accuracy of positive predictions and the completeness of recommended relevant items are important. Mathematically, F1 score can be defined as follows:
\begin{equation}
\text{F1} = 2 \cdot \frac{\text{Precision} \cdot \text{Recall}}{\text{Precision} + \text{Recall}} \nonumber
\end{equation}

Substituting the previously obtained precision and recall values gives \textbf{F1 score of  0.869}. This value demonstrates that our recommendation engine maintains a strong balance between giving relevant suggestions and minimizing missed ones.

\section{Discussion and Limitations}
The performance analysis shows that our content-based recommendation engine is effective in modeling individual user preferences and generating relevant suggestions. However, as a purely content-driven approach, it cannot capture collaborative patterns or emerging trends among similar users. The model also depends more on the quality and completeness of the metadata that are used to describe each video. Any noise or sparsity in these features can limit the accuracy of similarity computations. Moreover, content-based systems often lack serendipity, as they usually focus on what users already like instead of suggesting them to new but potentially relevant content.

To overcome these limitations, future work will explore hybridizing the system with collaborative filtering signals, which will use both personal and community-level patterns. We also plan to explore modern deep learning–based embedding techniques, such as BERT, and incorporate personalized TF–IDF weighting based on implicit user feedback to improve recommendation quality.

\section{Conclusion}
This paper presents the adaptation and implementation of a content-based recommendation engine for a video streaming platform. Unlike collaborative filtering, the content-based approach does not require data from other users which makes it particularly suitable for scenarios where personalized suggestions must be generated solely from an individual user's viewing and search history. As a result, content-based filtering proves to be an effective and fast approach for the video streaming platform. The preliminary experiments conducted with a selected group of user indicate that the proposed engine is capable of generating accurate and meaningful recommendations.

In the future, we will conduct experiments with a wider range of users having different tastes and viewing behaviors to further improve the accuracy of video predictions. Additionally, continuous research and development will be carried out on the engine to enhance the efficiency of the recommendation algorithm and provide a better user experience to the streaming platform’s users.

\bibliographystyle{IEEEtran}
\bibliography{ref}

\end{document}